%% file: 01_main.tex
\documentclass{article}
\usepackage{spconf}
\usepackage{amsmath,graphicx,hyperref}

\usepackage[utf8]{inputenc}
\usepackage{xcolor}
\usepackage{url}
\usepackage{graphicx}
\usepackage{amsmath}
\usepackage{subcaption}
\usepackage{rotating}
\usepackage{enumitem}
\usepackage{amssymb}
\usepackage[normalem]{ulem} % Load the ulem package

\definecolor{darkpastelgreen}{rgb}{0.01, 0.75, 0.24}
\definecolor{darkpink}{rgb}{0.91, 0.33, 0.5}
\apptocmd{\thebibliography}{\setlength{\itemsep}{0.9pt}\setlength{\parskip}{0.8pt}\linespread{0.9}\selectfont}{}{}
\title{ENHANCED GENERATIVE MACHINE LISTENER}

\name{Vishnu Raj, Gouthaman KV, Shiv Gehlot, Lars Villemoes and  Arijit Biswas}
\address{Dolby Laboratories}

\begin{document}

    \maketitle

    \begin{abstract}
        We present GMLv2, a reference-based model designed for the prediction of subjective audio quality as measured by MUSHRA scores.
        GMLv2 introduces a Beta distribution-based loss to model the listener ratings and incorporates additional neural audio coding (NAC) subjective datasets to extend its generalization and applicability. 
        Extensive evaluations on diverse testset demonstrate that proposed GMLv2 consistently outperforms widely used metrics, such as PEAQ and ViSQOL, both in terms of correlation with subjective scores and in reliably predicting these scores across diverse content types and codec configurations.
        Consequently, GMLv2 offers a scalable and automated framework for perceptual audio quality evaluation, poised to accelerate research and development in modern audio coding technologies.
    \end{abstract}

    \begin{keywords}
        Audio quality metrics, audio coding, deep learning, generative modeling
    \end{keywords}

    \input{10_introduction.tex}
    \input{30_method.tex}

    \input{70_results.tex}
    \input{90_concluding.tex}

    {
        \small
        \bibliographystyle{IEEEbib}
 	    \bibliography{99_library.bib}
    }

\end{document}

%% file: 10_introduction.tex
\section{Introduction}  \label{sec:intro}
The fundamental challenge in modern audio engineering is the accurate and scalable assessment of perceptual audio quality. Reliably quantifying the subjective human listening experience with algorithms is difficult due to the context-dependent nature of hearing, variations in acoustic content, and complex system-level distortions. The gold standard for evaluation remains subjective listening tests, standardized by protocols like ITU-R BS.1534 (MUSHRA)~\cite{bs20151534} and the BS.1116 series~\cite{bs1116}. However, these tests are prohibitively expensive, time-consuming, and impractical for real-time or large-scale applications. This creates a critical need for automated, objective frameworks that can not only predict audio quality accurately but also quantify the inherent uncertainty in perceptual judgments.  Consequently, there is a growing demand for automated, objective assessment frameworks that deliver perceptually aligned, sample-level quality predictions while remaining robust across diverse signal types, distortion categories, and listener populations.

Historically, this need has been met by reference-based (intrusive) methods that compare a degraded signal to its original version. Benchmark models such as PEAQ~\cite{thiede2000peaq}, PESQ~\cite{pesq}, and POLQA~\cite{polqa} became entrenched industry standards. 
These approaches were further advanced by methods like ViSQOL~\cite{visqol}, which improved perceptual alignment, and HASQI~\cite{hasqi} and HAAQI~\cite{haaqi}, which extended applicability to hearing-aid scenarios.
Meanwhile, reference-free deep learning models like NISQA~\cite{nisqa}, DNSMOS~\cite{dnsmos}, and Quality-Net~\cite{qualitynet} have introduced powerful alternatives without requiring reference signals. Despite these advances, intrusive methods remain the yardstick for high-precision benchmarking due to their strong grounding in signal comparison. However, a fundamental limitation across most prior approaches is their deterministic output: they predict a single scalar quality score, which inherently overlooks the variability and uncertainty inherent in human subjective evaluations. This limitation reduces their ability to capture the full complexity of listener perceptions, especially in ambiguous or edge-case audio scenarios where responses may vary widely.

The Generative Machine Listener (GMLv1) framework \cite{jiang2023generative} represented a key advance by modeling the full distribution of listener scores rather than a single point estimate, capturing perceptual uncertainty and delivering confidence intervals for more robust predictions. It achieved state-of-the-art performance across mono, stereo, and binaural audio quality tasks, highlighting the importance of modeling uncertainty in perceptual evaluation. However, GML uses Gaussian and Logistic likelihoods (unbounded and symmetric), which are not ideal for bounded MUSHRA scores, requiring artificial corrections that can distort shapes and calibration near boundaries, especially where listener data is asymmetric or truncated, emphasizing the need for a naturally bounded, flexible probabilistic model.

To address these shortcomings, we propose GMLv2, enhanced Generative Machine Listener, leveraging the Beta distribution. Naturally confined to the unit interval, the Beta distribution aligns perfectly with normalized MUSHRA scores, eliminating the need for post hoc adjustments. Its flexible shape parameters enable modeling of diverse distribution forms (including symmetric, skewed, and bimodal patterns) better capturing the complexity of listener scores. By using a neural network to predict the Beta distribution’s shape parameters 
$(\alpha, \beta)$, GMLv2 jointly estimates expected perceptual quality and uncertainty in a statistically principled way. This leads to better calibration and closer alignment with empirical data, providing a more reliable and interpretable framework for audio quality assessment. Accordingly, the main contributions of this work are as follows:

\begin{enumerate}

\item \textbf{Robust probabilistic framework}: GMLv2 introduces a statistically principled approach to audio quality assessment, providing predictions that are inherently aligned with listener data.

\item \textbf{Enhanced interpretability and reliability}: By jointly estimating perceptual quality and uncertainty, GMLv2 offers insights into both the expected scores and their uncertainty, supporting more informed evaluations.

\item \textbf{Improved generalization}: Flexibility of GMLv2 allows adaptation to heterogeneous audio content and diverse codec configurations, demonstrating consistent performance across a wide range of scenarios.
\end{enumerate}

%% file: 30_method.tex
\section{Generative Machine Listener-{v2}}   \label{sec:method}
Building on the probabilistic foundation established by~\cite{jiang2023generative}, GMLv2 predicts the parameters of a Beta distribution to capture both the expected perceptual quality and its uncertainty for a given pair of reference and degraded audio signals. 

\noindent
\textbf{Training Process.} 
Let $\textbf{x}$ be the reference signal and $\tilde{\textbf{x}}$ be the signal under test. Consider a model $\mathbf{y} = \mathbf{f}(\mathbf{x, \tilde{x}}; \mathbf{\theta})$ parameterized by $\mathbf{\theta}$ operating on the pair of inputs $(\mathbf{x, \tilde{x}})$ and predicts a vector $\mathbf{y}$ which characterizes the output distribution. The model is trained by minimizing the negative likelihood function $\mathcal{L}(\mathbf{x, \tilde{x}}, s) = - \log P(s; \mathbf{y})$, where $s$ denotes the individual listener MUSHRA score target.

\noindent
\textbf{Feature Extraction.} 
In order to model the auditory filter shape, \cite{de1968triggered} proposed `reverse correlation', in which the responses of a primary auditory nerve fiber to white noise stimuli are measured and correlated with the input. The `revcor' function thus obtained can be considered as an estimate of the impulse response of the peripheral auditory filter. The Gammatone Filter \cite{johannesma1972pre} is proposed as an analytic mathematical function that approximates measured revcor functions. The impulse response of Gammatone filter is defined as,
\begin{align}
    h(t) = \begin{cases}
        c t^{n-1} \exp(-2 \pi b t) \cos (2\pi f_0 t + \phi), & t >0 \\
        0 \qquad & t < 0
    \end{cases},
\end{align}
where $c$ is a proportionality constant, $n$ is the filter order, $b$ refers to the temporal decay coefficient with $b > 0$, $f_0$ is the center frequency of the filter and $\phi$ is the phase of the carrier wave.

We use Gammatone filter bank at the signal preprocessing stage as a frequency analysis component computing the Gammatone spectrogram, which in turn is fed into the deep neural architecture for prediction. We derive Left ($L$), Right ($R$), Mid ($M = (L + R) / 2$) and Side ($S = (L - R) / 2$) for both reference and test signal, compute Gammatone spectrogram (only power), concatenate the features and feed to the deep neural model.

\noindent
\textbf{Network architecture.}
The backbone is based on Inception blocks, wherein concatenated reference and degraded pairs are utilized as the network input \cite{biswas2022stereo,jiang2023generative} . 
Further,  the output layer is modified to accommodate the constraints of the parameters for Beta distribution.
Specifically, the final layer of the proposed architecture is a fully connected layer with 2-dimensional output. To preserve the unimodality of the predicted distribution, we need to have $\alpha, \beta > 1$. We achieve this by having $\mathbf{y} = [\tilde{\alpha}, \tilde{\beta}] = \mathbf{f}(\mathbf{x, \tilde{x}}; \mathbf{\theta})$ and compute the Beta distribution parameters as,
\begin{align}
    \alpha = 1 + \exp(\tilde{\alpha}), \qquad
    \beta = 1 + \exp(\tilde{\beta}).
    \label{eqn:beta-params}
\end{align}
We constrain the predicted Beta distribution to be unimodal (\ref{eqn:beta-params}) for aligning it with the unimodal nature of MUSHRA score distributions in our datasets.

\noindent
\textbf{Loss function.}
Noting that the actual MUSHRA scores are constrained between $0$ and $100$, we propose to use \emph{Beta} distribution to capture the output log-likelihood along with MUSHRA scores scaled to range $[0.0, 1.0]$. \emph{Beta} distribution is parameterized by two scalar parameters $(\alpha, \beta)$, and the density function is defined as
\begin{align}
    g(z; \alpha, \beta) = \frac{1}{B(\alpha, \beta)} z^{\alpha-1} (1 - z)^{\beta-1}, \label{eqn:beta_pdf}
\end{align}
where $\alpha > 0, \beta > 0, z \in [0, 1]$ and $B(\alpha, \beta)$ is the Beta function,
$
    \text{Beta}(\alpha, \beta) = \int \limits_{0}^{1} u^{\alpha - 1} (1 - u)^{\beta - 1} du. 
$
The mean of the distribution is defined as
    $\mu = \frac{\alpha}{\alpha + \beta},$    %\label{eqn:beta_mean}
and variance is defined as
    $Var[z] = \frac{\alpha \beta}{(\alpha + \beta)^2 (\alpha + \beta + 1)}.$  %\label{eqn:beta_var}
We can also characterize a unimodal Beta function with mode $\omega$ and concentration $\kappa$ with
\begin{align}
    \omega = \frac{\alpha -1}{\alpha + \beta - 2}, \qquad \kappa = \alpha + \beta.
\end{align}
Note that for $0 < \omega < 1$, we require $\alpha, \beta > 1$.

Let the training sample be $((\textbf{x}, \tilde{\textbf{x}}), s)$ and the model prediction can be written as 
$y = [\alpha, \beta] =  \mathbf{f}(\mathbf{x}_{GT}, \tilde{\mathbf{x}}_{GT}; \mathbf{\theta})$, 
where $\textbf{x}_{GT}$ and $\tilde{\textbf{x}}_{GT}$ refers to the Gammatone spectrogram of reference and test signal respectively. The loss function, used to train the model, can be derived as,

\begin{align}
    \small \mathcal{L}(\mathbf{x, \tilde{x}}, s) = -(\alpha - 1) \ln s - (\beta - 1) \ln (1 - s) + \ln B(\alpha, \beta). \label{eqn:beta_loss}
\end{align}

\noindent
\textbf{Inference Process.}
For a given pair of reference and test signals, the audio quality of the latter is quantified by simultaneously predicting \emph{MUSHRA} scores and \emph{confidence intervals} (CIs)

\noindent \textit{i) MUSHRA Prediction:} We first predict the estimated parameters for the Beta distribution $(\hat{\alpha}, \hat{\beta})$ using the trained model. The corresponding MUSHRA score is then computed as:
\begin{align}
    \text{Predicted MUSHRA} = \frac{\hat{\alpha}}{\hat{\alpha} + \hat{\beta}} \times 100.   \label{eqn:pred_mushra}
\end{align}
\textit{ii) Confidence Intervals Prediction:} CI concerns the estimation of a true source parameter through repeated experiments. Since the model directly furnishes the distribution parameters, we follow a more pragmatic approach to calculate CIs ~\cite{jiang2023generative}. Conventionally, the simulation of a test with $N$ listeners begins by drawing  $N$ samples from the predicted output distribution, which are then analyzed using the same procedures as in the original listening tests. In our approach, we deviate slightly from this procedure by using the model output directly to represent the mean, as defined in (\ref{eqn:pred_mushra}), and by scaling the Beta distribution variance $(\hat{\alpha}, \hat{\beta})$ by $100^2$ to represent the score variance. To maintain consistency with the test sets, confidence intervals are computed using the $t$-distribution with estimated mean and variance. However, the quantitative evaluation of CI prediction is left for future work

%% file: 70_results.tex
% Results

\begin{table*}[!th]
\small
\scalebox{0.885}{
\centering
\begin{tabular}{l|rr|rr|rrr|rrr|rrr}
\hline
\multicolumn{1}{c|}{} &
  \multicolumn{2}{c|}{\textbf{PEAQ} \cite{thiede2000peaq}} &
  \multicolumn{2}{c|}{\textbf{ViSQOL} \cite{visqol}} &
  \multicolumn{3}{c|}{\textbf{GMLv1} \cite{jiang2023generative}} &
  \multicolumn{3}{c|}{\textbf{GMLv1*}\cite{jiang2023generative}} &
  \multicolumn{3}{c}{\textbf{Proposed GMLv2}} \\ \hline
\multicolumn{1}{c|}{\textbf{Eval. set}} &
  \multicolumn{1}{c}{$R_p^\uparrow$} &
  \multicolumn{1}{c|}{$R_s^\uparrow$} &
  \multicolumn{1}{c}{$R_p^\uparrow$} &
  \multicolumn{1}{c|}{$R_s^\uparrow$} &
  \multicolumn{1}{c}{$R_p^\uparrow$} &
  \multicolumn{1}{c}{$R_s^\uparrow$} &
  \multicolumn{1}{c|}{$\text{OR}^\downarrow$} &
%   \multicolumn{1}{c}{\textbf{RMSE}} &
  \multicolumn{1}{c}{$R_p^\uparrow$} &
  \multicolumn{1}{c}{$R_s^\uparrow$} &
  \multicolumn{1}{c|}{$\text{OR}^\downarrow$} &
%   \multicolumn{1}{c}{\textbf{RMSE}} &
  \multicolumn{1}{c}{$R_p^\uparrow$} &
  \multicolumn{1}{c}{$R_s^\uparrow$} &
  \multicolumn{1}{c}{$\text{OR}^\downarrow$}
%   \multicolumn{1}{c}{\textbf{RMSE}} 
  \\ \hline
\textbf{USAC-1}
   % PEAQ
   & $0.4700$
   & $0.4000$
   % ViSQOL
   & $0.8090$
   & $0.8356$
   % GMLv1
   & $0.8788$
   & $0.8779$
   & $0.8333$
%    & $0.9144$
   % GMLv1+
   & $0.9137$
   & $\mathbf{0.8985}$
   & $0.8333$
%    & $2.0107$
   % GMLv2
   & $\mathbf{0.9191}$
   & $0.8964$
   & $\mathbf{0.0451}$ 
%    & $0.8699$ \\
\\
\textbf{USAC-2}
   % PEAQ
   & $0.4200$
   & $0.2000$
   % ViSQOL
   & $0.7712$
   & $0.7775$
   % GMLv1
   & $0.8905$
   & $0.8574$
   & $0.7167$
%    & $1.2247$
   % GMLv1+
   & $0.8925$
   & $0.8426$
   & $0.7250$
%    & $2.7600$
   % GMLv2
   & $\mathbf{0.9261}$
   & $\mathbf{0.8919}$
   & $\mathbf{0.0667}$
%    & $1.0266$
\\
\textbf{USAC-3} 
   % PEAQ
   & $0.5600$
   & $0.6200$
   % VisQOL
   & $0.8229$
   & $0.9036$
   % GMLv1
   & $0.9233$
   & $0.9443$
   & $0.5947$
%    & $1.8088$
   % GMLv1+
   & $0.9218$
   & $0.9170$
   & $0.7992$
%    & $3.2532$
   % GMLv2
   & $\mathbf{0.9442}$
   & $\mathbf{0.9348}$
   & $\mathbf{0.0455}$
%    & $1.4281$
\\
\textbf{Binaural 1}
   % PEAQ
   & $0.7500$
   & $0.7900$
   % ViSQOL
   & $0.8976$
   & $0.9271$
   % GMLv1
   & $0.9827$
   & $\mathbf{0.9544}$
   & $0.2987$
%    & $4.0574$
   % GMLv1+
   & $0.9480$
   & $0.9330$
   & $0.5584$
%    & $4.8838$
   % GMLv2
   & $\mathbf{0.9840}$
   & $0.9418$
   & $\mathbf{0.1818}$
%    & $2.5682$
\\
\textbf{Binaural 2}
   % PEAQ
   & $0.4200$
   & $0.5600$
   % ViSQOL
   & $0.9613$
   & $0.8539$
   % GMLv1
   & $0.9830$
   & $0.8968$
   & $0.3214$
%    & $2.5573$
   % GMLv1+
   & $0.9768$
   & $0.9064$
   & $0.6429$
%    & $1.7450$
   % GMLv2
   & $\mathbf{0.9895}$
   & $\mathbf{0.9068}$
   & $\mathbf{0.0119}$
%    & $1.5476$
\\
\textbf{NAC Mono}
   % PEAQ
   & $0.3400$
   & $0.3100$
   % ViSQOL
   & $0.8920$
   & $0.8614$
   % GMLv1
   & $0.9514$
   & $0.9210$
   & $0.4643$
%    & $4.2844$
   % GMLv1+
   & $0.9211$
   & $0.9386$
   & $0.8333$
%    & $5.2358$
   % GMLv2
   & $\mathbf{0.9706}$
   & $\mathbf{0.9398}$
   & $\mathbf{0.0714}$
%    & $2.8523$
\\
\textbf{NAC Stereo}
   % PEAQ
   & $0.5800$
   & $0.4000$
   % ViSQOL
   & $0.8157$
   & $0.8948$
   % GMLv1
   & $0.8689$
   & $0.8175$
   & $0.5222$
%    & $3.7598$
   % GMLv1+
   & $0.9283$
   & $0.8991$
   & $0.5889$
%    & $4.5564$
   % GMLv2
   & $\mathbf{0.9492}$
   & $\mathbf{0.9254}$
   & $\mathbf{0.0778}$
%    & $2.1208$
\\
\textbf{ODAQ}
   % PEAQ
   & $0.7100$
   & $0.6500$
   % ViSQOL
   & $0.6995$
   & $0.7954$
   % GMLv1
   & $0.7092$
   & $0.7537$
   & $0.7833$
%    & $2.0$
   % GMLv1+
   & $0.8127$
   & $0.8095$
   & $0.8167$
%    & $4.6539$
   % GMLv2
   & $\mathbf{0.8256}$
   & $\mathbf{0.8287}$
   & $\mathbf{0.2708}$
%    & $2.7235$
\\ \hline
% \textbf{Average}
%    % PEAQ
%    & $0.5313$
%    & $0.4913$
%    % ViSQOL
%    & $0.8337$
%    & $0.8562$
%    % GMLv1
%    & $0.8985$
%    & $0.8779$
%    & $0.5668$
% %    & $2.0$
%    % GMLv1+
%    & $0.9144$
%    & $0.8931$
%    & $0.7247$
% %    & $4.6539$
%    % GMLv2
%    & \shortstack{$\mathbf{0.9385}$ \\ $\textcolor{darkpastelgreen}{\scriptstyle \blacktriangle 4.01\%}$}
%    & \shortstack{$\mathbf{0.9082}$ \\ $\textcolor{darkpastelgreen}{\scriptstyle \blacktriangle 1.51\%}$}
%    & \shortstack{$\mathbf{0.0964}$ \\  $\textcolor{darkpastelgreen}{\scriptstyle \blacktriangle -47.05\%}$}
% %    & $2.7235$
% \\ \hline
\textbf{Agg. Score}
   % PEAQ
   & $0.5606$
   & $0.5187$
   % ViSQOL
   & $0.8468$
   & $0.8643$
   % GMLv1
   & $0.9234$
   & $0.9006$
   & $0.5668$
%    & $2.0$
   % GMLv1+
   & $0.9284$
   & $0.9038$
   & $0.7247$
%    & $4.6539$
   % GMLv2
   & \shortstack{$\mathbf{0.9526}$ \\ $\textcolor{darkpastelgreen}{\scriptstyle \blacktriangle 0.0292}$}
   & \shortstack{$\mathbf{0.9205}$ \\ $\textcolor{darkpastelgreen}{\scriptstyle \blacktriangle 0.0199}$}
   & \shortstack{$\mathbf{0.0964}$ \\  $\textcolor{darkpastelgreen}{\scriptstyle \blacktriangle -0.4704}$}
%    & $2.7235$
\\ \hline
\end{tabular}}
\vspace{-1em}
\caption{Comparative analysis of proposed GMLv2 framework with existing methods on diverse evaluation sets. The proposed method achieves consistently higher correlations with subjective ratings and lower outlier ratios than prior models, demonstrating robust generalization across diverse codecs, channel configurations, and content types. Best results are highlighted in bold. The $\textcolor{darkpastelgreen}{\scriptstyle \blacktriangle}$ represents improvement with proposed GMLv2 as compared to existing GMLv1. The $*$ denotes the GMLv1 trained with revised dataset (Sec.~\ref{sec:results}). Aggregate score is computed using Fisher's Z-transform approach as discussed in \cite{torcoli2021objective}.}
\label{tab:results}
\end{table*}

\section{Experiments and Results}   \label{sec:results}
\subsection{Experimental Setup}
\noindent
\textbf{Training set} consists of MUSHRA scores from a variety of traditional codecs covering stereo and binaural audio, as well as neural audio codecs (NACs) at different bitrates. MUSHRA listening tests were conducted with $48$~kHz audio using headphones~\cite{jiang2023generative,biswas2025rf}. For stereo codecs, the set includes different content types encoded with AAC~\cite{AAC}, HE-AAC v1/v2~\cite{den2009overview}, and Dolby AC-4~\cite{riedmiller2017delivering} across various bitrates. For binaural tests, the set contains Dolby Atmos~\cite{robinson2012scalable} encoded with AC-4 Immersive Stereo (IMS)~\cite{dolby2021ac4}, Dolby Digital Plus Joint Object Coding (DD+JOC)~\cite{purnhagen2016immersive}, and AC-4 Advanced Joint Object Coding (A-JOC)~\cite{riedmiller2017delivering}, with the Dolby Atmos binaural rendition serving as the unencoded reference. In addition, we include the binauralized 3GPP IVAS codec~\cite{bruhn20253gpp,multrus2024immersive} for ambisonic signals. For NACs, the mono and stereo sets include Encodec~\cite{defossez2022highfi}, Descript Audio Codec~\cite{kumar2023high}, and MDCTNet~\cite{villemoes2024mdctnet} samples at different bitrates, covering a wide range of content. This set also incorporates neural enhancements of traditional codecs using GANs~\cite{DCAE} and SampleRNN~\cite{RoySampleRNNAudio}. In total, our dataset consists of $82{,}191$ sample pairs, including $68{,}503$ traditional codec samples and $14{,}688$ neural codec samples. For training, we randomly split the dataset into $90\%$ training and $10\%$ validation.

\noindent
\textbf{Evaluation set}. To benchmark the performance of the proposed method, we used the test sets described in \cite{jiang2023generative} (including USAC tests \cite{quackenbush2011performance,quackenbush2013mpeg} and binaural tests), along with samples from the ODAQ dataset \cite{torcoli2024odaq} and in-house listening test samples collected using NACs for mono and stereo settings. A detailed description is given below:

\begin{enumerate}
   \setlength{\itemsep}{1pt}
   \setlength{\parskip}{0pt}
   \setlength{\parsep}{0pt}
%\begin{itemize}[noitemsep,nolistsep]
% \small
    \item \emph{USAC-1} \cite{quackenbush2011performance}. $27$ mono audio items, scored by $66$ listeners, resulting in $288$ paired data points.
    \item \emph{USAC-2} \cite{quackenbush2011performance}. $27$ stereo audio items encoded at lower bitrates, scored by $44$ listeners, resulting in $240$ paired data points.
    \item \emph{USAC-3} \cite{quackenbush2011performance}. $27$ stereo audio items encoded at higher bitrates, scored by $28$ listeners, resulting in $264$ paired data points.
    \item \emph{Binaural-1} \cite{jiang2023generative}. $11$ binaural renderings at high bitrates, scored by $9$ listeners, resulting in $77$ paired data points. 
    \item \emph{Binaural-2} \cite{jiang2023generative}. $12$ binaural renderings at low bitrates, scored by $11$ listeners, resulting in $84$ paired data points. 
    \item \emph{NAC mono}. $12$ mono audio items were encoded with MDCTNet (48 kHz) at 24 kb/s 
    (demo content from~\cite{villemoes2024mdctnet}) 
    and compared with the publicly available 16 kb/s DAC model (44.1 kHz)~\cite{Descript_16kbps}, coded at 16, 12, and 8 kb/s in a MUSHRA test. Ratings from $12$ listeners yielded $84$ paired data points.
    \item \emph{NAC stereo}. $15$ stereo audio items from the ODAQ set~\cite{torcoli2024odaq}, encoded using publicly available 16 kb/s DAC model at 16 and 32 kb/s (with left and right channels encoded independently), were compared against 48 kHz EnCodec stereo model~\cite{defossez2022highfi} at 24 kb/s. Ratings from $13$ listeners yielded $90$ paired data points.
    \item \emph{ODAQ} \cite{torcoli2024odaq}. $25$ audio samples processed by $6$ methods at $5$ different quality levels, rated by $26$ expert listeners, resulting in $240$ samples.
% \end{itemize}
\end{enumerate}

\begin{figure*}
    \centering
    \begin{subfigure}{0.24\textwidth}
        \includegraphics[width=\textwidth]{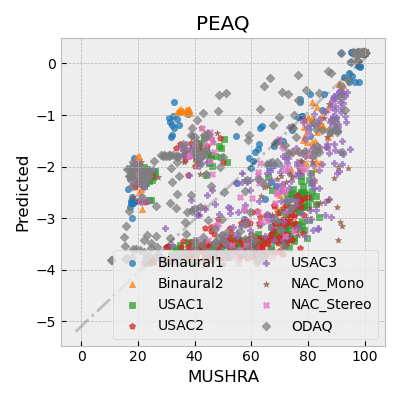}
        \label{fig:scatter_peaq}
    \end{subfigure}
    \hfill
    \begin{subfigure}{0.24\textwidth}
        \includegraphics[width=\textwidth]{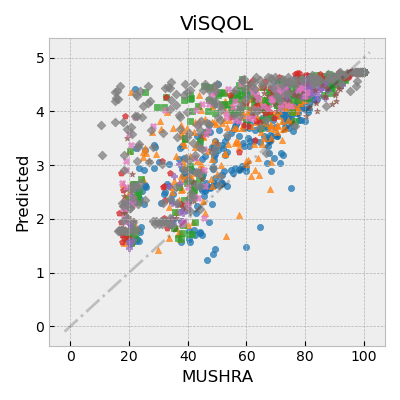}
        \label{fig:scatter_visqol}
    \end{subfigure}
    \hfill
    \begin{subfigure}{0.24\textwidth}
        \includegraphics[width=\textwidth]{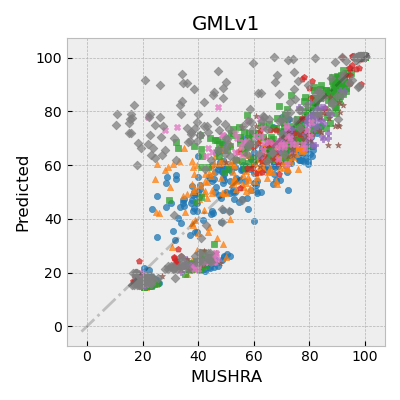}
        \label{fig:scatter_gmlv1}
    \end{subfigure}
    \hfill
    \begin{subfigure}{0.24\textwidth}
        \includegraphics[width=\textwidth]{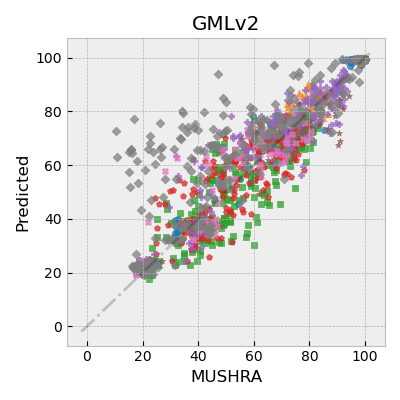}
        \label{fig:scatter_gmlv2}
    \end{subfigure}
    \vspace{-2em}
    \caption{Visual comparison of predicted versus ground-truth MUSHRA scores shows that the proposed model closely aligns with the identity line across all test sets, with minimal outliers, demonstrating substantially higher predictive accuracy and robustness than PEAQ and ViSQOL.}
    \label{fig:scatter}
    \vspace*{-0.3cm}
\end{figure*}

\noindent\textbf{Training configuration.}
The proposed method is implemented in PyTorch v2.1 and trained with 2 NVIDIA A10G GPUs. Spectograms use a 80ms window, 20ms hop, 32 channels, and 50 Hz as the lowest frequency. Training employed a batch size of $8$/GPU, Adam optimizer with a learning rate of $1 \times 10^{-4}$, $400{,}000$ steps. Models are evaluated using Pearson ($R_p$) and Spearman ($R_s$) correlations and Outlier ratio (OR) at $95\%$ CI \cite{yi2022conferencingspeech}, with the best checkpoint chosen to maximize $R_p \times R_s$ on the validation set.

\subsection{Results}
We provide comparisons of the proposed approach against three popular methods: (i) PEAQ (Perceptual Evaluation of Audio Quality)~\cite{thiede2000peaq,tsp_peaq}, (ii) ViSQOL-v3 (Virtual Speech Quality Objective Listener)~\cite{chinen2020visqol} in general audio mode, and (iii) GMLv1~\cite{jiang2023generative}. PEAQ is designed for evaluating high-fidelity general audio and predicts the Objective Difference Grade (ODG) for the signal under test. ViSQOL-v3 predicts MOS-LQO (Mean Opinion Score -- Listening Quality Objective). Finally, GML, which forms the basis of this work, is also designed for general audio and predicts MUSHRA test scores for the signal under test. In addition to the proposed architectural and loss-function changes, we also use NAC MUSHRA tests to train our model. To ensure a fair comparison, we retrained GMLv1~\cite{jiang2023generative} with the new data and compared it against the proposed approach.

A quantitative comparison (Table~\ref{tab:results}) of audio quality models reveals that the proposed method significantly outperforms all prior work across eight test sets. Baseline models like PEAQ show the lowest correlation, struggling with modern codecs, while ViSQOL offers improvements but remains inconsistent on complex stereo content. The GML model surpasses these, demonstrating high correlation but suffering from a high Outlier Ratio (OR), indicating frequent perceptual mismatches, a problem that worsens even when its correlation is boosted with new training data. In stark contrast, the proposed model achieves the highest and most consistent performance across all metrics, delivering 
excellent 
correlation ($R_p > 0.92$) and, most importantly, the lowest OR in nearly all test sets. This robust behavior, which holds across diverse codecs (including unseen ones), channel configurations, and content types, affirms the model's superior generalization and its ability to accurately predict human perceptual ratings with significantly fewer errors.
While the proposed framework captures uncertainty through the concentration of the predicted Beta distribution, quantitative validation of Confidence Intervals is reserved for future work.

Figure~\ref{fig:scatter}
presents a visual comparison of predicted versus ground-truth scores across test sets for the methods under evaluation. In the figures, the x-axis represents the ground-truth MUSHRA values, and the y-axis represents the predictions from each model. 
The plots show that predictions from PEAQ are widely scattered and deviate significantly from the ideal, indicating poor performance. While ViSQOL presents a substantial improvement with tighter clustering, it still shows inconsistencies on more complex stereo (NAC Stereo) and artifact-heavy (ODAQ) audio. In contrast, the proposed model's predictions provide the best agreement with the ground-truth identity line across all datasets, demonstrating a strong linear relationship with minimal outliers. This tight clustering, even on the most challenging test sets, visually reinforces the quantitative findings that the proposed model offers significantly better predictive accuracy and robustness than competing methods like PEAQ and ViSQOL.

%% file: 90_concluding.tex
\section{Concluding Remarks}    \label{sec:concluding}
We proposed a novel reference-based perceptual audio quality metric. Building on \cite{jiang2023generative}, we introduced a Beta loss as a principled approach to train MUSHRA score predictions. By incorporating additional quality measurement datasets for NACs, in addition to those used in \cite{jiang2023generative}, we extended the scope of GMLv2 to support NAC use cases. Our extensive experiments, covering both traditional codecs and NACs, demonstrate that GMLv2 outperforms most popular audio quality evaluation metrics in the majority of cases. Finally, through systematic analysis of predicted scores, we showed that the proposed model provides the best-calibrated results across the spectrum of audio content and codec types.